\documentclass[pdflatex,iicol,sn-mathphys-num]{sn-jnl}

\usepackage{graphicx}
\usepackage{booktabs}
\usepackage{tabularx}
\usepackage{amsmath,amssymb,amsfonts}
\usepackage{xcolor}
\usepackage{textcomp}
\usepackage{placeins}
\usepackage{caption}
\usepackage{needspace}
\usepackage{url}

\begin{document}

\title{Teams Without Teams: A Trace-Based Method for Measuring Team-Like Organization}

\author[1,2]{\fnm{Thomas Baillie-Dawe} \sur{}}\email{Thomas.Baillie-Dawe@imt-atlantique.fr}
\author[1]{\fnm{Cécile Bothorel} \sur{}}\email{Cecile.Bothorel@imt-atlantique.fr}
\author[2]{\fnm{Nicolas Jullien} \sur{}}\email{Nicolas.Jullien@imt-atlantique.fr}

\affil[1]{\orgname{Lab-STICC UMR CNRS 6285, IMT Atlantique}, \orgaddress{\city{Brest}, \country{France}}}
\affil[2]{\orgname{LEGO-Marsouin, IMT Atlantique}, \orgaddress{\city{Brest}, \country{France}}}



\abstract{
Digital platforms provide increasingly detailed traces of collaborative activity, yet these traces do not directly reveal whether contributions remain largely independent or whether more structured forms of collective work emerge through interaction. This paper addresses the methodological problem of measuring such organization when the collective being studied is not defined as a team in advance. We propose a trace-based method for measuring \textit{teamness}, understood as the degree to which collaborative activity expresses properties associated with teamwork. The method translates established dimensions of teamwork into behavioral indicators that can be recovered from digital traces. These indicators can be examined separately and summarized through a composite measure of their joint expression. It does not require predefined team membership and is designed to be adapted across collaborative settings. We illustrate the method using revision histories from 52,437 articles in the English Wikipedia's Medicine WikiProject, comprising more than 10 million human-authored edits. Using article-months as observation units, the application shows that different dimensions of team-like organization display distinct patterns and that their joint expression can vary within the same articles over time. The contribution is a general measurement framework for examining graded and dynamic forms of collective organization directly from behavioral traces.

}


\maketitle

\section{Introduction}

Digital platforms provide increasingly detailed records of collaborative activity, making it possible to observe how people work together at a scale that is difficult to capture through surveys or direct observation alone. Yet detailed traces do not tell us directly whether collaborative activity remains largely independent or whether more structured forms of collective work emerge through interaction. The methodological challenge is therefore to develop measures that can quantify the degree of collective organization expressed in these behavioral traces \citep{lazerComputationalSocialScience2009}.

Team research provides one theoretical basis for doing so. It identifies recurring  properties of organized collective work associated with team-like forms of organization. Coordination, interdependence, role differentiation and boundedness are recurring features used to characterize teams and teamwork \citep{kozlowskiWorkGroupsTeams,cohenWhatMakesTeams1997,ilgenTeamsOrganizationsInputProcessOutput2005,marksTemporallyBasedFramework2001}. Importantly, observing these properties does not require us to assume that the contributors involved constitute a team. They can instead be used to assess the extent to which a collective episode displays features associated with teamwork.

Empirically, however, these properties are generally measured after the relevant team has already been identified. Researchers begin with a known collaborative unit and then examine its structures, processes, efficiency or outcomes. Research on fluid, dynamic, ad hoc and temporary teams has relaxed assumptions of stable membership, but the collaborative unit under study generally remains identifiable \citep{grossman2024fluidteams,driskellFluidVirtualTeams2023,funkeEditorialFluidTeams2024}.

Open online collaboration presents a different measurement problem. Contributors can enter and leave a collaborative task without formal membership, and the set of participants may change substantially over time. Their activity is observable, but there may be no predefined team from which measurement can begin. The question is therefore not simply whether a team exists, but to what extent observed collaborative activity displays properties associated with teamwork.

The contribution of this paper is a trace-based method for answering this question. Building on existing work that conceptualizes \textit{teamness} as graded and multidimensional \citep{cookeTeamsTeamnessFuture2024}, we translate established dimensions of teamwork into behavioral indicators that can be recovered from digital traces. These indicators can be examined separately and summarized through a composite measure of their joint
expression. The method does not require team membership to be specified in advance: it compares collaborative episodes according to the degree to which their observed activity expresses team-like properties. The method is defined independently of a particular platform so that its operationalization can be adapted to different collaborative environments.

We illustrate the method using revision histories from the English Wikipedia's Medicine WikiProject. Wikipedia provides a particularly useful setting because collaboration is organized around shared articles while participation remains open and fluid. Treating all contributors to an article as a team would impose a social unit before examining how their activity is organized. Instead, we use article-months as observation units and measure variation in team-like organization directly from revision traces. The empirical analysis is exploratory and is intended to demonstrate the type of collaborative structure that the method makes observable.

The contribution of the paper is therefore methodological: a trace-based method for measuring graded team-like organization from behavioral traces. It enables collaborative episodes to be compared and changes in their organization to be followed over time.

The article is organized as follows. Section~\ref{sec:litrev} establishes the conceptual and measurement foundations of the approach. Section~\ref{sec:method} presents the trace-based method, which constitutes the core methodological contribution of the paper. Section~\ref{sec:data} illustrates the method using Wikipedia. Section~\ref{sec:results} presents the exploratory results, and Section~\ref{discussion} discusses the implications of the approach and directions for future research.

\section{Related Work and Conceptual Foundations} 
\label{sec:litrev}

This section establishes the conceptual and measurement foundations of the approach. We first identify the core features of teamwork that provide the basis for measurement, before examining the difficulty of applying these features when the collective is not defined as a team in advance. We then use teamness as a graded property and show why this perspective requires a measurement approach capable of recovering variation in team-like organization directly from collaborative activity.

\subsection{Core features of teamwork}
\label{sec:features}

To study teamwork empirically, it is necessary to identify the core structural features that characterize team-like collaboration. Across fields such as organizational psychology, management, sociology or computer-supported cooperative work, several recurring dimensions appear in definitions of teamwork. Teams are generally understood as groups of individuals who coordinate their actions, rely on one another's contributions and work toward a shared objective over a period of time \citep{kozlowskiWorkGroupsTeams,cohenWhatMakesTeams1997,hackmanLeadingTeamsSetting2002,ilgenTeamsOrganizationsInputProcessOutput2005}. This requires interdependence between members, role differentiation, coordination of activities, and some degree of bounded membership \citep{marksTemporallyBasedFramework2001,maloneInterdisciplinaryStudyCoordination1994}. These dimensions do not capture the full complexity of teamwork, but they provide a functional baseline that appears consistently across the literature. In this paper, we treat these elements as the minimal structural features that characterize team-like collaboration.

\textit{Shared purpose} is the defining element that distinguishes teams from collections of individuals acting in parallel \citep{kozlowskiWorkGroupsTeams,cohenWhatMakesTeams1997,hackmanLeadingTeamsSetting2002}. In formal organizational settings, these goals are often explicitly stated, prescribed through hierarchical authority, or embedded in task design and performance systems \citep{hendersonStrategicAlignmentLeveraging1999,wagemanTeamDiagnosticSurvey2005,KatesGalbraith2007_DesigningYourOrganization,klojcnikProjectManagementSystematic,lockeBuildingPracticallyUseful2002,lathamNewDevelopmentsDirections2007}. Shared purpose is mainly assessed through team members' self-reports of whether goals are shared and commonly understood across the team \citep{wagemanTeamDiagnosticSurvey2005}.

A second core feature is \textit{interdependence}. Teams are not simply groups of co-present individuals; they are groups in which members depend on one another to complete the work \citep{hertelManagingDistanceInterdependence2004,berntzenAutonomousInterdependentRoles2021}. In formal team settings, interdependence is often designed into the task itself through sequential or tightly coupled workflows and routines \citep{puranamWhatsNewNew2014,courtrightStructuralInterdependenceTeams2015,raveendranRoleInterdependenceMicrofoundations2020}. Interdependence has, for example, been measured through team members' perceptions of their dependence on one another \citep{courtrightStructuralInterdependenceTeams2015}. In online collaboration, this dependence can instead arise through the work itself, as contributors build on and modify one another's contributions \citep{farajKnowledgeCollaborationOnline2011,majchrzakContradictoryInfluenceSocial2013}.

A third feature is \textit{role differentiation}. Team members do not all do the same thing. Division of labor and repeated specialization are central to many definitions of teamwork because they reduce redundancy and allow complementary contributions to emerge \citep{ilgenTeamsOrganizationsInputProcessOutput2005,marksTemporallyBasedFramework2001}. Roles may be formally assigned, but they can also emerge through repeated patterns of behavior. In organizational research, role structure has been examined through formal role assignments, member reports, and observational coding. Research on online interaction has further shown that emergent roles can be identified from recurring behavioral and structural patterns \citep{welser2007,welserFindingSocialRoles2011,arazyTurbulentStabilityEmergent2016}.

A fourth feature is \textit{coordination}. Coordination refers to the way teammates manage dependencies so that individual actions fit together into a coherent whole \citep{maloneInterdisciplinaryStudyCoordination1994}. Classic work shows that coordination mechanisms vary with the degree and form of interdependence, ranging from routines and standardization to more continuous mutual adjustment \citep{mathieuInfluenceSharedMental,kotlarskyCoordinatingExpertiseKnowledge2014}. Coordination has traditionally been studied through surveys and direct observation of team interaction, but online collaboration has also led researchers to infer it from behavioral traces. In Wikipedia, Kittur and Kraut (2008) for example use the concentration of editing activity as a proxy for implicit coordination: when a smaller set of contributors accounts for more of the work, coordination demands are expected to be lower \citep{kitturHarnessingWisdomCrowds2008}.

Finally, teamwork usually assumes some form of \textit{boundedness}. Even if teams vary in stability, researchers generally assume that it is possible to identify who is participating in the collective effort over a given period of time. In most traditional approaches, membership remains a basic analytical starting point \citep{hackmanLeadingTeamsSetting2002,ilgenTeamsOrganizationsInputProcessOutput2005}. Boundedness has therefore typically been assessed through membership rosters, agreement about who belongs to the team, diagnostic questions concerning team membership, and characteristics such as team size \citep{wagemanTeamDiagnosticSurvey2005}.


Taken together, these dimensions provide an established basis for characterizing teamwork. Their empirical measurement, however, has generally begun with a collective that is already identified. This makes it difficult to examine how strongly team-like properties are expressed when no team is predefined, without simply equating an observable platform-level unit, such as all contributors to an article, with a team. Open collaboration therefore requires a way to assess team-like organization without assuming the social unit that the measurement is intended to characterize.

\subsection{When the team is not predefined}

The organizational conditions under which teams are traditionally studied do not map onto open online collaboration. However, the absence of predefined teams does not imply the absence of forms of organization associated with teamwork. Contributors can coordinate their actions through a shared artifact rather than through explicit group membership \citep{omahonyEmergenceGovernanceOpen2007,marshStigmergicEpistemologyStigmergic2008}. Interdependence can emerge as contributors build on or repair one another's work \citep{majchrzakContradictoryInfluenceSocial2013}, while role differentiation can emerge behaviorally without formal assignment \citep{welser2007,welserFindingSocialRoles2011}, and the shared artifact can also provide a common object of work around which contributors orient their activity \citep{clark1996,Hutchins1995_CognitionInTheWild,heylighenStigmergyUniversalCoordination2016}.

Open online collaboration leaves detailed behavioral traces of how contributors work together, while team research provides established concepts for describing features associated with teamwork. The difficulty lies in connecting these two levels: translating established concepts of teamwork into measures that can be recovered reliably from behavioral traces.

At the same time, the study of teams has increasingly moved beyond the assumption of stable, long-term membership. Research on fluid, dynamic, temporary, and ad hoc teams shows that teamwork can occur despite changing membership and limited prior familiarity \citep{ilgenTeamsOrganizationsInputProcessOutput2005,mathieuTeamEffectiveness199720072008,lacerenzaMissingTeamDynamics2025,driskellFluidVirtualTeams2023,funkeEditorialFluidTeams2024}. These approaches show that stability and familiarity are not necessary conditions for teamwork. Yet even in these approaches, the collaborative unit is generally identifiable before its properties are measured. Membership may change, sometimes considerably, but these changes occur within the frame of a defined team or task \citep{hirstEffectsMembershipChange2009,mortensenConstructingTeamAntecedents2014}.

Open collaboration therefore requires measures that can capture variation in the organization of collaborative activity itself. Rather than treating team-like organization as an all-or-nothing property, the aim is to assess how strongly its constituent features are expressed in observed behavior. This problem extends beyond Wikipedia to other digitally mediated settings.

\subsection{Teamness as a graded property} 

The idea that a collection of individuals can constitute a group to a greater or lesser degree is not new. Campbell's concept of \textit{entitativity} introduced groupness as a graded property, referring to the extent to which an aggregate is perceived as a coherent social entity rather than as a simple collection of individuals \citep{campbellCommonFateSimilarity1958}.

This perspective was subsequently developed by Hamilton and Sherman and by Lickel et al., who examined the properties associated with stronger perceptions of groupness \citep{hamiltonPerceivingPersonsGroups1996,lickelVarietiesGroupsPerception2000}. Lickel et al. (2000), in particular, show that interaction, shared goals and shared outcomes are among the properties most strongly associated with perceived entitativity.

Entitativity, however, has primarily been developed and applied as a perceptual construct \citep{blanchardDevelopingEntitativityMeasure2020, lickelVarietiesGroupsPerception2000}. Empirical studies generally present respondents with a specified aggregate and ask them either to rate how much it constitutes a group or to evaluate properties associated with entitativity. The aggregate being evaluated is therefore specified in advance. Entitativity establishes that groupness can vary in degree, but its dominant measurement tradition does not provide a way to assess this variation directly from behavioral traces. 

A similar move toward graded rather than categorical distinctions has developed within team research. Hollenbeck, Beersma, and Schouten (2012), for example, argue that the proliferation of categorical team types can be replaced by continuous dimensions along which teams vary, including authority differentiation and temporal stability \citep{hollenbeckBeyondTeamTypes2012}.

More recently, Cooke et al. (2024) explicitly propose moving ``from teams to teamness,'' treating teamness as multidimensional rather than requiring a strict distinction between teams and non-teams \citep{cookeTeamsTeamnessFuture2024}.
Under this perspective, collaborating aggregates can express different degrees and configurations of team-like properties. We build on this existing use of \textit{teamness} rather than introducing it as a new theoretical construct. This graded perspective also allows the degree of teamness to vary over time \citep{marksTemporallyBasedFramework2001}. Research on team dynamics has long shown that team processes and structures are not static, while work on teaming emphasizes collective activity as something that unfolds through interaction rather than as a fixed property of a stable group
\citep{edmondsonTeamingOrganizationsLearn2012}.

In a related tradition, Brubaker (2004) argues that groupness should be treated as a variable and, at times, as an event rather than assuming the continuous existence of bounded groups \citep{brubakerEthnicityGroups2004}.
Team-like organization can therefore be treated as multidimensional and time-varying. The same collaborative setting may display stronger or weaker expressions of the features of teamwork at different points in time. Yet a graded conception of teamness does not by itself resolve the measurement problem. Entitativity measures rely largely on member or observer ratings of a named aggregate \citep{blanchardDevelopingEntitativityMeasure2020}, while team diagnostics and emerging teamness measures generally begin with an identifiable team or collaborative ensemble and draw on surveys, self-reports, or experimental tasks \citep{wagemanTeamDiagnosticSurvey2005}.

These approaches can assess how group-like or team-like a specified collective is, but they do not provide a way to measure team-like organization directly from the behavioral traces of naturally occurring collaborative episodes. This distinction is important: the aim is not to identify teams as a predefined latent category, but to compare observed collections of contributors according to the extent to which their activity expresses team-like properties.

Existing team-based measures provide limited leverage for comparing the full continuum from collections displaying little team-like organization to collections displaying stronger combinations of teamwork properties. Survey- and rating-based approaches also make it difficult to identify when such organization emerges and dissolves, to observe structured collaboration that occurs without explicit awareness of belonging to a team, and to follow changes in teamness repeatedly within the same collaborative system. Studying teamness in open collaboration therefore requires a measurement approach that does not depend on prior team designation or member perception and that can capture variation in team-like organization directly from observable behavior over time.

\section{Trace-Based Method for Measuring Team-Like Organization
}
\label{sec:method}

This section presents the trace-based method that constitutes the methodological contribution of the paper. It defines how team-like properties can be recovered from behavioral traces independently of a specific platform. The following section applies this method to Wikipedia.

\subsection{Applicability requirements}

The method is designed for collaborative settings in which activity can be observed through digital traces. Its application requires three elements.

\begingroup
\emergencystretch=2em

\begin{enumerate}
\renewcommand{\labelenumi}{(\theenumi)}

\item First, collaboration must be organized around an identifiable object, task, or unit that provides a common scope for activity. Depending on the setting, this may be a Wikipedia article or project, a software repository or issue, an OpenStreetMap task, a shared document, or a work item in a project-management system.

\item Second, individual actions must be attributable to actors and timestamped, so that their sequence and patterns of interaction can be reconstructed.

\item Third, the traces must contain sufficient information to operationalize the relevant dimensions of teamwork, such as dependencies between actions or differences in the types of contributions performed.

\end{enumerate}

\endgroup

These requirements do not depend on whether a formal team already exists. The method can be used to compare predefined teams, but also to examine collections of contributors for which team membership is not known in advance. The specific traces used to satisfy these requirements depend on the collaborative environment. Section~\ref{sec:data} shows how they are instantiated using Wikipedia revision histories.


\subsection{Data primitives}\label{sec:primitives}

We represent collaborative activity as a sequence of recorded actions $e$. Each action is characterized by five primitives: an actor $u(e)$, a timestamp $t(e)$, the collaborative object or task $p(e)$ to which the action belongs, a dependency relation $par(e)$ linking it to the prior contribution or state on which it acts and an action type $r(e)$ describing its functional category.

These primitives provide the information needed to operationalize the dimensions of teamness. Actor and timestamp information captures the temporal organization of activity used to assess coordination. Dependency relations identify when one actor builds on the contribution of another and provide the basis for measuring interdependence. Action types describe how work is distributed across actors and are used to assess role differentiation. Finally, the set of actors participating within a given collaborative scope provides the basis for assessing boundedness. Section~\ref{sec:data} specifies how these primitives are applied in Wikipedia revision data.

\subsection{Indicators}\label{sec:indicators}

The data primitives are used to derive behavioral indicators corresponding to the dimensions of teamwork established in Section~\ref{sec:features}. Shared purpose is incorporated through the definition of the collaborative scope: the actions considered within an observation unit are oriented toward a common object. Coordination, interdependence, role differentiation, and boundedness are then assessed from the organization of activity within that scope. The indicators are defined independently of any particular platform; application-specific parameters are introduced in Section~\ref{sec:data}.

\paragraph{Coordination.}
Coordination is captured through temporal responsiveness between actors. This follows the conception of coordination as mutual adjustment, in which actors adapt their activity in relation to the actions of others \citep{maloneInterdisciplinaryStudyCoordination1994,mathieuInfluenceSharedMental}.

Let $E=\{e_1,\ldots,e_k\}$ denote the set of actions in an observation unit, ordered by time. For each action $e_i$, let $u(e_i)$ denote the actor who performed it and $t(e_i)$ its timestamp. Let $\tau_r$ denote the duration of the response window used to identify rapid responses. The value of $\tau_r$ is application-specific and should reflect the temporal rhythm of interaction in the setting under study. It should therefore be reported explicitly and its sensitivity to alternative plausible windows assessed.

\begin{equation}
\label{eq:fastreply}
\begin{aligned}
F(e_i) = \mathbb{1}\big[\,\exists\, e_j \in E,\ j<i :\;&
u(e_j) \neq u(e_i) \\
&\land\; t(e_i)-t(e_j) \leq \tau_r
\,\big].
\end{aligned}
\end{equation}

The coordination score is the share of actions that satisfy this condition within an observation unit containing $k$ actions:

\begin{equation}\label{eq:coord}
\text{Coordination Share (Coord)}
=
\frac{\sum_{i=1}^{k} F(e_i)}{k}.
\end{equation}

Temporal co-activity provides a complementary measure of coordination by capturing periods in which several actors are active within the same time window. The observation period is divided into consecutive intervals $I_j$ of duration $\tau_c$. An interval $I_j$ is considered co-active when it contains actions from more than one actor:

\begin{equation}\label{eq:coactivity_window}
C(I_j) =
\mathbb{1}\left[
\left|\{u(e): e \in I_j\}\right| > 1
\right].
\end{equation}

Let $\mathcal{I}^{+}$ denote the set of active intervals, that is, intervals containing at least one action. The co-activity score is the proportion of active intervals involving more than one actor:

\begin{equation}\label{eq:coac}
\text{Co-activity Share (Co-ac)}
=
\frac{
\sum_{I_j \in \mathcal{I}^{+}} C(I_j)
}{
|\mathcal{I}^{+}|
}.
\end{equation}


The distribution of rapid responses across actor pairs can additionally be used to characterize the structure of coordination. Each qualifying pair of actions contributes one rapid-response interaction; when several earlier actions satisfy the response-window condition for the same later action, each qualifying interaction is retained.

Let $G_m$ denote the share of rapid-response interactions accounted for by the $m$ most active interaction pairs.

\begin{equation}\label{eq:backbone}
B = \text{Coord} \times G_m.
\end{equation}

When no rapid-response interaction is observed, we define $B=0$.

Higher values indicate observation units in which rapid cross-actor interaction is both frequent and concentrated within a recurrent interaction backbone. The value of $m$ is application-specific.

\paragraph{Interdependence.}
Interdependence captures the extent to which the actions of one actor depend on contributions made by another. This corresponds to task coupling, in which one member's output becomes an input into another's work \citep{courtrightStructuralInterdependenceTeams2015,raveendranRoleInterdependenceMicrofoundations2020,hertelManagingDistanceInterdependence2004}. In artifact-mediated collaboration, such dependencies can be observed when actors build on or modify previous contributions \citep{majchrzakContradictoryInfluenceSocial2013}.

For an action $e_i$ with dependency relation $par(e_i)$, a cross-actor dependency is defined as:

\begin{equation}\label{eq:interdep}
D(e_i) =
\mathbb{1}\left[
par(e_i)\text{ is observed}
\;\text{and}\;
u(e_i) \neq u(par(e_i))
\right].
\end{equation}

The share of cross-actor dependencies within an observation unit provides a measure of interdependence:

\begin{equation}\label{eq:interdep_share}
\text{Interdependence Share (D)}
=
\frac{\sum_{i=1}^{k}D(e_i)}
{k}.
\end{equation}


The direction of these dependencies provides an additional indication of their structure. Reciprocity captures whether dependency relationships between pairs of actors are one-way or mutual. Let $A_{uv}=1$ when actor $u$ performs at least one action whose parent contribution was authored by actor $v$, and $A_{uv}=0$ otherwise. Let $\mathcal{P}$ denote the set of unordered actor pairs $\{u,v\}$ for which at least one dependency is observed in either direction. A pair is reciprocal when $A_{uv}=A_{vu}=1$.

\begin{equation}\label{eq:reciprocity}
R =
\frac{
\left|\left\{\{u,v\}\in\mathcal{P}: A_{uv}=1 \text{ and } A_{vu}=1\right\}\right|
}{
|\mathcal{P}|
}.
\end{equation}

Reciprocity is defined only for observation units in which at least one cross-actor dependency relationship is observed.

\paragraph{Role differentiation.}
We operationalize role differentiation through functional specialization in the distribution of work. Rather than requiring formally assigned roles, the method examines whether actors systematically concentrate their activity in particular functional categories. This builds on work showing that differentiated roles can emerge from recurring patterns of behavior in online collaboration \citep{welser2007,welserFindingSocialRoles2011}.

Let $Q$ denote the number of action categories and $x_{u,r}$ the number of actions of type $r$ performed by actor $u$. The action categories are defined according to the functional forms of work observable in the setting under study and must be specified as part of the application. Actor-level specialization is measured as the share of the actor's activity belonging to their most frequent action category:

\begin{equation}\label{eq:spec}
\mathrm{Spec}(u)
=
\max_{r \in \{1,\ldots,Q\}}
\left(
\frac{x_{u,r}}
{\sum_{q=1}^{Q}x_{u,q}}
\right).
\end{equation}

By construction, $\mathrm{Spec}(u)$ ranges from $1/Q$ to $1$, with higher values indicating that an actor concentrates more of their observed activity in a single functional category. Let $U$ denote the set of active actors in the observation unit and $n=|U|$. Role differentiation at the level of the observation unit is summarized by the average specialization of its active actors:

\begin{equation}\label{eq:specshare}
\text{Specialization Share (S)}
=
\frac{1}{n}
\sum_{u \in U} \mathrm{Spec}(u).
\end{equation}

\paragraph{Boundedness.}
Boundedness concerns the scope of participation within which the other dimensions are assessed. In settings without formal membership, the method does not require a predefined roster. Instead, the relevant collective is defined by the set of actors active within the collaborative scope and observation period. Applications may impose additional activity or participation requirements when needed to obtain interpretable indicators, but these are implementation choices rather than part of the generic definition of teamness.

\subsection{Composite representation of teamness}\label{sec:measure} 

The indicators provide a multidimensional representation of team-like organization and can be examined separately. When a single measure is useful, they can additionally be combined into a composite teamness score $T$. Because the indicators may differ in scale and may also vary with the overall volume of activity, they are first standardized among observations with comparable activity levels. %

\begin{equation}\label{eq:index}
T = \sum_{j=1}^{J} w_j z_j,
\end{equation}

where $J$ is the number of components included in the composite, $z_j$ is the standardized score for component $j$, and $w_j$ is its weight.

Higher values of $T$ indicate a stronger joint expression of the behavioral properties included in the composite. The score is continuous and is used to compare degrees of team-like organization rather than to classify observations as teams or non-teams. Component scores should be retained alongside $T$ so that different configurations of teamness remain observable. When a composite is used, its sensitivity to the weighting scheme should also be assessed.

\section{Application to Wikipedia}
\label{sec:data}

\subsection{Data source}

The empirical corpus is derived from the English Wikipedia's Medicine WikiProject. This collaborative initiative focuses on articles related to medical and health topics and provides a large corpus of collaborative revision histories.

The dataset contains 52,437 unique articles and initially comprised 11,465,869 revisions produced by 1,948,961 unique editors. After removing edits generated by 1,464 automated bot accounts (representing 1,039,407 revisions), the final corpus contains 10,426,462 human-authored edits.

\begin{figure*}[h]
    \centering
    \includegraphics[width=0.9\linewidth]{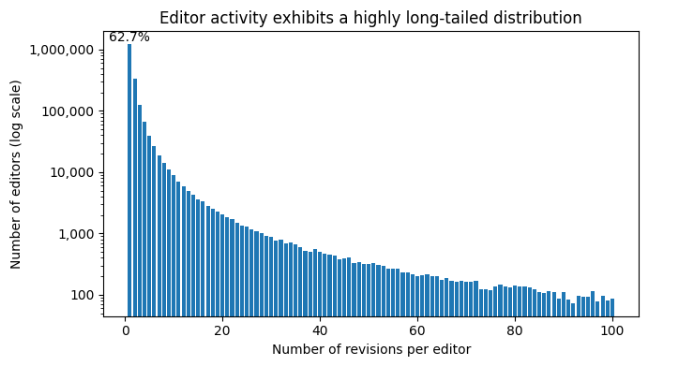}
    \caption{Distribution of editor contributions in WikiProject Medicine}
    \label{fig:editor_distribution}
\end{figure*}

Analysis of participation patterns reveals the highly skewed contribution structure typical of large-scale online collaboration systems. As shown in Figure \ref{fig:editor_distribution}, editorial participation follows a long-tail distribution in which a large majority of users contribute only once or a few times, while a small minority of contributors produce the majority of edits.

Specifically, around two-thirds of all unique editors contributed only a single revision. More than 90\% of editors made five or fewer contributions, while fewer than 10\% contributed more than five times. This pattern highlights the presence of a large peripheral population alongside a smaller set of more active contributors.

This structure is also reflected at the article level. Many articles are maintained by relatively small groups of recurring contributors rather than by broad collective participation.

The dataset spans articles created between 2001 and 2024.

\begin{figure*}[h]
    \centering
    \includegraphics[width=0.9\linewidth]{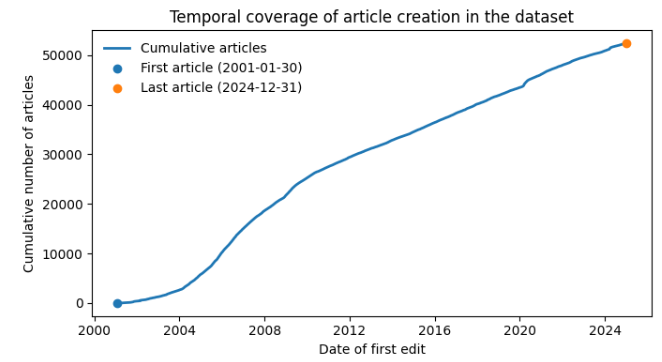}
    \caption{Temporal evolution of article creation in WikiProject Medicine}
    \label{fig:article_creation}
\end{figure*}

As shown in Figure \ref{fig:article_creation}, article creation accelerated rapidly during Wikipedia's early expansion, peaking between 2005 and 2006. Most of the article corpus was established by 2010, after which article creation continued at a slower but steady pace.

\subsection{Observation units}

The continuous revision history of each article was segmented into \textit{article-month} observation units. Each unit captures non-revert, human edits made to a given article during a given calendar month. We denote the set of edits observed for article $p$ in month $m$ as $E_{p,m}=\{e_1,\ldots,e_k\}$.

This operationalization reflects our theoretical perspective that teamwork in open collaboration should be studied as temporary episodes of activity rather than as stable social groups. The monthly interval provides a balance between temporal granularity and sufficient interaction density. Shorter intervals would produce sparse observations, while longer intervals would smooth over collaboration dynamics.

Prior to constructing the indicators, three filtering steps were applied. First, the dataset was restricted to the article namespace, that is, data corresponding to encyclopedia articles, excluding discussion pages, user profiles and administrative content. Second, all edits generated by automated bots were removed. Third, revert actions were excluded. While revert activity is an important component of Wikipedia governance and may itself reflect collaborative processes, we exclude it here to focus the operationalization on constructive revision activity. The component-level analysis focuses on active article-months containing at least 11 non-revert human edits, corresponding to the activity bands used in the analysis below.

\subsection{Operationalization on Wikipedia}

In the Wikipedia application, an action $e$ corresponds to a non-revert human revision, the actor $u(e)$ to its editor, and the collaborative scope $p(e)$ to the article being edited. Revision timestamps provide $t(e)$, parent revision identifiers provide the dependency relation $par(e)$, and edit types provide the functional categories $r(e)$ used to assess specialization.

For coordination, we set the rapid-response window to $\tau_r=6$ hours. An edit is classified as a rapid response when at least one earlier edit by another editor occurred on the same article within the preceding six hours and within the same article-month. We additionally calculate the share of rapid interactions accounted for by the three most active interaction pairs and combine this with the coordination share to obtain the coordination-backbone component $B$ defined in Equation~\ref{eq:backbone}.

We evaluated response windows of 1, 3, 6, 12, and 24 hours and examined how the coordination measure changed across these specifications. The six-hour window was retained for the Wikipedia application based on this sensitivity analysis.

For co-activity, we set $\tau_c=1$ hour and divide the observation period into calendar hours. An active hour is considered co-active when at least two editors contribute during that hour, and the co-activity score is the proportion of active hours satisfying this condition.

Interdependence is derived from parent--child revision relations. A revision is cross-editor dependent when its parent revision can be identified and was authored by another editor. Parent revisions are resolved across the revision history and may therefore precede the current article-month. Reciprocity is calculated at the level of editor dyads: a dyad is reciprocal when cross-editor dependencies are observed in both directions. Interdependence share $D$ is reported descriptively, while reciprocity $R$ is used as the interdependence component of the composite index.

Functional specialization is operationalized using three edit-type categories ($Q=3$): reference-related edits, substantial additions, and other edits. Reference-related edits are identified when the edit summary contains citation-related terms such as ``ref'', ``cite'', ``doi'', ``pmid'', or ``isbn''. Among the remaining edits, those adding at least 200 bytes are classified as substantial additions, while all other edits are assigned to the residual category. For each editor, specialization is the share of their edits belonging to their dominant category, and article-month specialization is the mean of these editor-level scores.

\subsection{Composite teamness index}

For the composite illustration, we restrict the analysis to article-months containing sufficient behavioral information to estimate all four composite components on a common basis. We therefore retain article-months with 3--15 active editors, at least 20 edits available for estimating specialization, at least 10 observed dependency pairs, and at least 10 active hours, with all four component measures observed. These thresholds are operational rather than theoretical boundaries: some cutoff is necessary to distinguish observations containing sufficient trace information from those in which one or more components are only weakly supported. Applying these requirements yields 10,529 article-months for the composite analysis.

For the Wikipedia illustration, the four composite components are the coordination-backbone score, co-activity, reciprocity, and functional specialization. Each component is standardized within activity bands defined by monthly edit volume: mid-activity (11--30 edits), high-activity (31--100 edits), and extreme-activity ($>100$ edits). We evaluated alternative activity thresholds and examined how the resulting indicator distributions changed across different band definitions. The 11, 31, and 100-edit cutoffs were retained because they provided a stable separation between activity levels while preserving sufficient observations within each band. Means and standard deviations are calculated within each activity band, pooling article-months across articles and years. The resulting $z$-scores therefore express each article-month relative to observations with comparable editing activity.

We summarize their joint expression using the following weighted specification:

\begin{equation}
\label{eq:coeff-wikipedia}
T = 0.35z_B + 0.25z_\text{Co-ac} + 0.20z_R + 0.20z_\text{Spec}.
\end{equation}

We  assess the sensitivity of the composite to the weighting scheme by recomputing it with equal weights across all four components. The weighted and equal-weight indices are highly correlated across activity bands and produce very similar patterns, indicating limited sensitivity to the precise weighting scheme.

\begin{table}[!htbp]
\centering

\caption{Sensitivity of the composite teamness index to equal weighting. $T_w$ denotes the weighted specification and $T_{eq}$ the equal-weight specification.}

\label{tab:teamness_robustness}
\begin{tabular}{lccc}
\toprule
Activity band & Corr. & Median $T_w$ & Median $T_{eq}$ \\
\midrule
Mid     & 0.980 & -0.078 & -0.066 \\
High    & 0.974 & -0.064 & -0.037 \\
Extreme & 0.967 & -0.059 & -0.032 \\
\bottomrule
\end{tabular}
\end{table}

In the empirical analysis, we use $T$ as an exploratory summary of the joint expression of these components rather than as a definitive classification of teams.

\section{Exploratory Results}\label{sec:results}

The Wikipedia application is intended to illustrate what the trace-based method makes observable when applied to naturally occurring collaborative activity, rather than to provide an exhaustive characterization of collaboration on Wikipedia. We focus on two descriptive aspects of the application: variation in the component indicators across levels of editing activity and variation in their joint expression within the same collaborative objects over time.

First, the component indicators do not vary uniformly with editing activity. Using mid-activity article-months as a descriptive reference, rapid responses increase from approximately 10\% of edits in this group to approximately 45\% in extreme-activity months. Cross-editor interdependence varies considerably less across the same activity bands, remaining within approximately 24--36\% of edits. These comparisons illustrate that the indicators capture different aspects of collaborative organization rather than simply reproducing the same pattern of activity.

Second, the composite measure makes it possible to examine how the joint expression of the component indicators varies within the same article over time. Because the components of $T$ are standardized within activity bands, the composite is interpreted relative to article-months with comparable editing activity rather than as an absolute scale across activity levels.

\begin{figure}[!htbp]
    \centering
    \includegraphics[width=\linewidth]{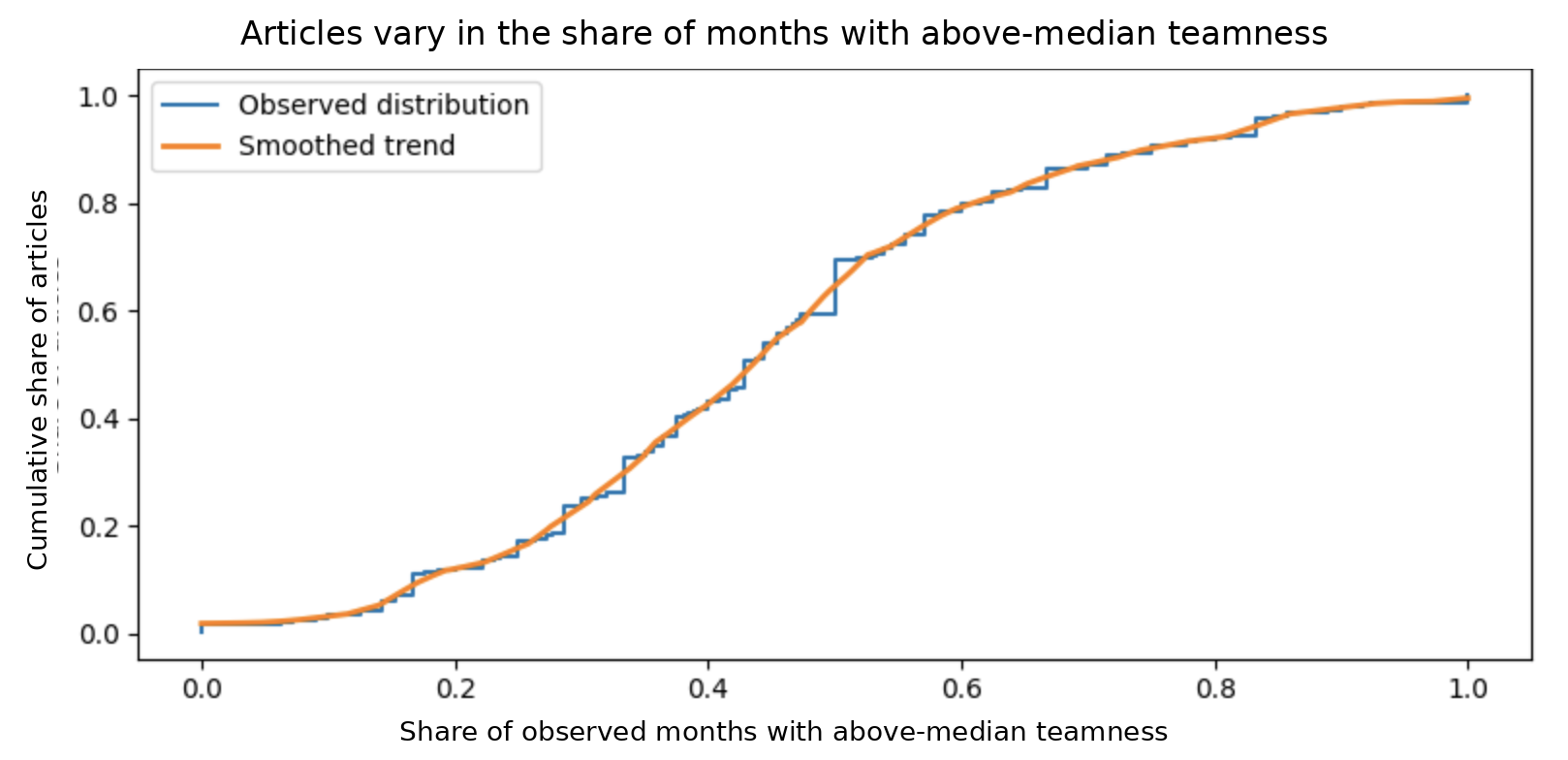}
    \caption{Cumulative distribution of the share of observed months with above-median composite teamness, among articles with at least six composite observations.}
    \label{fig:team-like-organization}
\end{figure}

Figure~\ref{fig:team-like-organization} shows, for articles represented by at least six observations in the composite analysis, the share of observed months in which $T$ falls above the median of its corresponding activity band. Variation in this share shows that articles differ in how consistently their observed months fall in the upper half of the teamness distribution. The same collaborative object therefore need not be characterized by a single, fixed level of team-like organization.

Taken together, these exploratory results illustrate two uses of the trace-based method. The component indicators distinguish different aspects of collaborative activity, while their joint representation allows variation in collective organization to be followed within the same collaborative setting over time.

\section{Discussion and Conclusion}\label{discussion}

This paper develops a trace-based method for assessing the extent to which collaborative activity expresses properties associated with teamwork. Building on established work that conceptualizes teamness as graded and multidimensional \citep{cookeTeamsTeamnessFuture2024}, the approach operationalizes these properties directly from behavioral traces.

The contribution is methodological. Existing work provides the conceptual basis for treating teamness as a matter of degree; we provide a way to operationalize this perspective from behavioral traces. The method translates dimensions of teamwork into observable indicators that can be examined separately and, when useful, summarized through a composite measure. Rather than requiring a team to be designated before measurement, the organization expressed in collaborative activity itself becomes the object of analysis.

The Wikipedia application illustrates two consequences of this approach. First, the component indicators need not display the same relationship with overall activity. Temporal responsiveness and cross-editor dependency, for example, reveal different patterns across article-months. Second, combining the indicators makes it possible to compare the same collaborative object across time. The variation observed within articles illustrates that team-like organization need not be treated as a fixed property.

The application also highlights the distinction between the general measurement framework and its implementation in a particular setting. The framework specifies the trace information required to recover different dimensions of team-like organization, while the concrete operationalizations depend on the collaborative environment. In Wikipedia, temporal response windows, parent--child revision relations, and functional edit categories provide observable proxies for coordination, interdependence, and specialization. These measures capture particular behavioral manifestations of the underlying dimensions rather than exhaustive representations of them. The composite score similarly provides one way of summarizing their joint expression and should be interpreted alongside its component indicators. Because functional specialization is estimated from observed edit categories, scores for contributors with very few actions may be mechanically high. The concentration component of the coordination-backbone measure is also partly constrained by the number of active editor pairs, since concentration among the three most active pairs becomes mechanically higher when few pairs are available.

Several directions follow from this work. The trace-based indicators should first be validated against independent measures of team organization, connecting behavioral measurement to established survey, observer-based, or experimental approaches. Richer traces, including discussion activity and more detailed classifications of contribution types, could extend the collaborative processes captured by the framework. Relating the indicators to outcomes such as article quality would also make it possible to examine when team-like organization is consequential for collective performance. Finally, applying the method to other collaborative settings would allow the operationalizations and composite specification to be tested across different forms of digitally mediated work.

The paper therefore contributes a trace-based framework for measuring graded forms of team-like organization from behavioral activity. Its central purpose is to provide a way of comparing how strongly different properties associated with teamwork are expressed and how their joint expression changes over time.

\backmatter

\section*{Declarations}

\paragraph{Funding}
This work was supported by a French government grant managed by the Agence Nationale de la Recherche as part of the France 2030 program, references 22-EXEN-0005 (PEPR eNSEMBLE / CONGRATS) and 24-EXEN-0002 (Data2Laws project).

\paragraph{Competing interests}
The authors have no competing interests to declare.

\paragraph{Data availability}
The data used in this study are derived from publicly available Wikipedia revision histories. Processed data supporting the findings of this study are available from the corresponding author upon reasonable request.

\bibliography{sn-bibliography}

\end{document}